\documentclass[conference]{IEEEtran}
\IEEEoverridecommandlockouts
\usepackage{times}
\usepackage{amsmath}
\usepackage{amssymb}
\usepackage{multicol,multirow}
\usepackage[tableposition=top]{caption}
\usepackage{cite}
\usepackage{booktabs}
\usepackage{bigstrut}
\usepackage{floatrow}
\usepackage{rotating}
\usepackage{adjustbox}
\usepackage{amsmath}

\def\BibTeX{{\rm B\kern-.05em{\sc i\kern-.025em b}\kern-.08em
    T\kern-.1667em\lower.7ex\hbox{E}\kern-.125emX}}
\begin{document}
\title{Feature Enhancer Segmentation Network (FES-Net) for Vessel Segmentation}


\author{Tariq M. Khan,~\IEEEmembership{Member,~IEEE,}
  Muhammad Arsalan,~\IEEEmembership{Member,~IEEE,}
  Shahzaib Iqbal,~\IEEEmembership{Member,~IEEE,}\\
        Imran Razzak,~\IEEEmembership{Member,~IEEE,}
        and~Erik Meijering,~\IEEEmembership{Fellow,~IEEE}
\thanks{Tariq M. Khan, Imran Razzak and Erik Meijering are with the School of Computer Science \& Engineering, UNSW, Sydney, Australia (e-mail: \{tariq.khan, imran.razzak, erik.meijering\}@unsw.edu.au)}  
\thanks{Shahzaib Iqbal is with the Department of Electrical and Computer Engineering, COMSATS University Islamabad (CUI), Islamabad, Pakistan}
\thanks{Muhammad Arsalan is with the Department of Computer Science, Qatar University, Doha, Qatar}
}

\maketitle

\begin{abstract}
    
Diseases such as diabetic retinopathy and age-related macular degeneration pose a significant risk to vision, highlighting the importance of precise segmentation of retinal vessels for the tracking and diagnosis of progression. However, existing vessel segmentation methods that heavily rely on encoder-decoder structures struggle to capture contextual information about retinal vessel configurations, leading to challenges in reconciling semantic disparities between encoder and decoder features. To address this, we propose a novel feature enhancement segmentation network (FES-Net) that achieves accurate pixel-wise segmentation without requiring additional image enhancement steps. FES-Net directly processes the input image and utilizes four prompt convolutional blocks (PCBs) during downsampling, complemented by a shallow upsampling approach to generate a binary mask for each class. We evaluate the performance of FES-Net on four publicly available state-of-the-art datasets: DRIVE, STARE, CHASE, and HRF. The evaluation results clearly demonstrate the superior performance of FES-Net compared to other competitive approaches documented in the existing literature.
\end{abstract}

\vspace{0.5\baselineskip}



\section{Introduction}

\IEEEPARstart{C}{omputer}-aided wide-scale screening presents a feasible approach to the detection of diseases in people with diabetes mellitus, as it has the potential to increase scarce healthcare resources and support healthcare professionals \cite{Fraz2012,imtiaz2021screening}. Previous research \cite{crosby2012retinal,habib2014association,naqvi2019automatic} has underscored the importance of the size and structure of the retinal vessels in the diagnosis and prediction of the prognosis of diabetic retinopathy. The noticeable alterations in vessel dimensions act as robust markers of the severity of the disease and could be used to forecast the potential future progression of the condition, as supported by the conclusions of these studies \cite{crosby2012retinal}.

In computer-aided diagnosis (CAD) systems, segmentation of retinal blood vessels is crucial and time-consuming \cite{naveed2021towards}. This is due to the fact that many retinal abnormalities, such as hemorrhages and microaneurysms, typical manifestations of diabetic retinopathy, are often observed close to these vessels \cite{iqbal2022recent}. Furthermore, efficient segmentation of retinal vessels can allow estimation of the topology of vessel maps \cite{Zhao2018,khan2023retinal}. 


Segmenting retinal blood vessels is a challenging task in retinal image analysis due to the presence of numerous obstacles. These include low contrast, uneven intensity, and varying thickness between primary vessels and capillaries present in images of the retinal fundus. Additionally, the presence of exudates and lesions further complicates the segmentation process. To address these challenges, numerous studies have employed supervised or unsupervised algorithms and computer vision techniques, in order to achieve accurate and automatic segmentation \cite{iqbal2023robust,iqbal2023ldmres,Soomro2018,Khawaja2019a,khan2018robust}. The latest research points out that deep learning architectures have better performance compared to other techniques \cite{Soomro2019,khan2020region,khan2022t}.

Retinal vessel segmentation has been aided by a number of approaches based on deep learning. U-Net is proposed for medical image segmentation, but it identifies false boundaries in retinal images along with blood vessels \cite{Ronneberger2015}. Gu \textit{et al.} put forth a context encoder network that captures higher-order information while maintaining spatial data \cite{Gu2019CENetCE} for the segmentation of blood vessels. Yan \textit{et al.} enhanced U-Net's performance by introducing a joint loss function \cite{yan2018joint}.

Wang \textit{et al.} introduced DEU-Net, which uses a fusion module function to merge a spatial path with a large kernel, thus preserving spatial data while capturing semantic details \cite{Wang2019a}. Fu \textit{et al.} presented DeepVessel, which uses a CNN featuring a multiscale and multilevel layer to construct a dense hierarchical representation \cite{fu2016deepvessel}. Additionally, they incorporated a conditional random field to model extended interactions among pixels \cite{fu2016deepvessel}. Notwithstanding the effectiveness of these approaches, they overlook the computational commutation needed to customize the network for use with manageable resources on embedded systems.

\begin{figure*}[!t]
  \centering
  \includegraphics[width=1\textwidth]{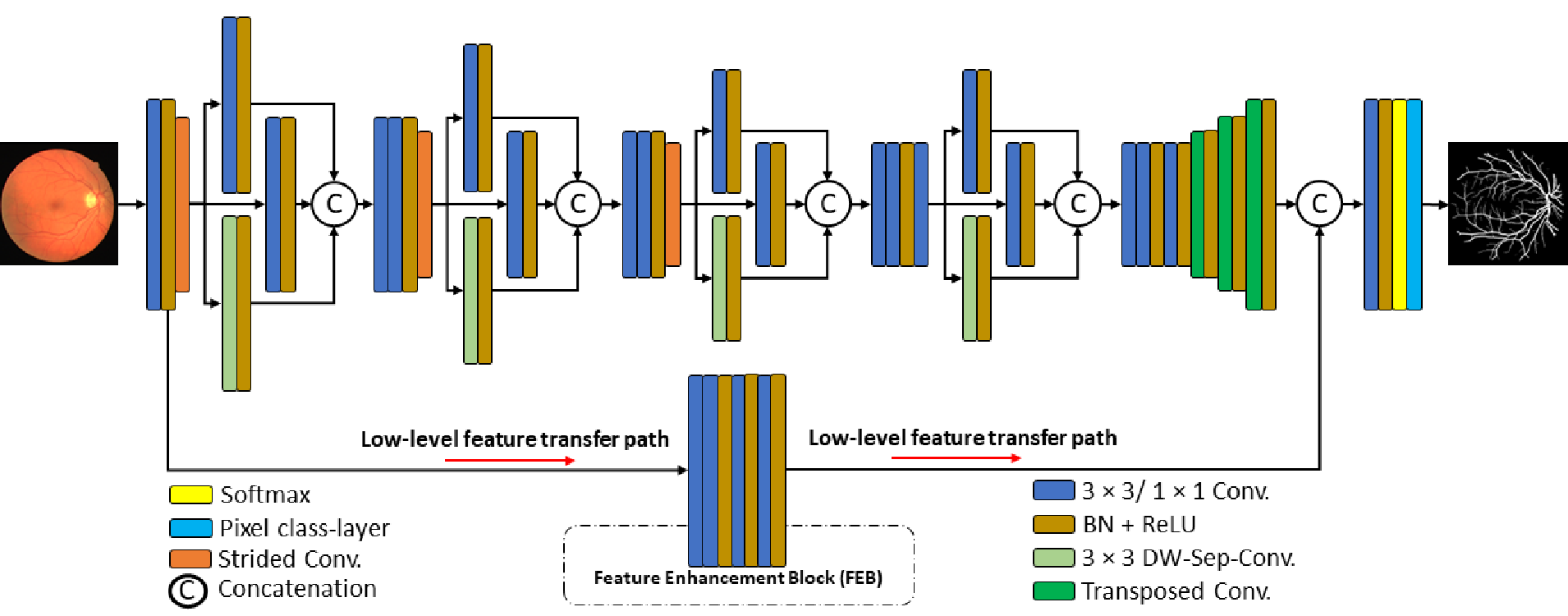}
  \caption{Architecture of the proposed FES-Net. BN = batch normalization. DW-Sep-Conv = depth-wise separable convolution. FEB = feature enhancement block. ReLU = rectified linear unit activation.}
  \label{FES-NET}
\end{figure*}

\begin{figure}[!t]
  \centering
  \includegraphics[width=0.7\textwidth]{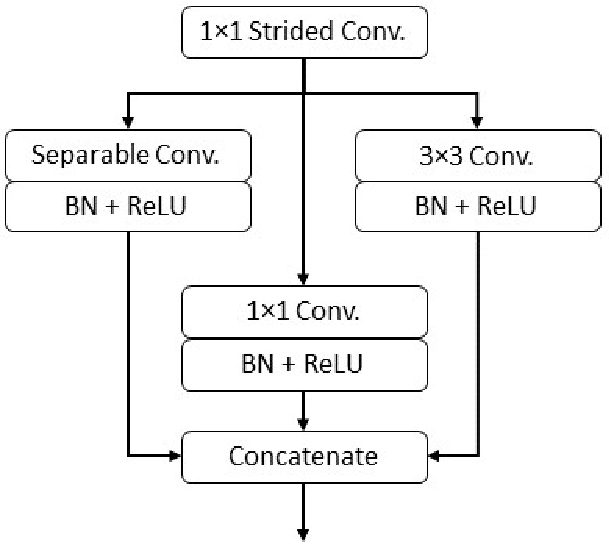}
  \caption{Connectivity schematic of prompt convolutional blocks (PCBs) used in the proposed FES-Net.}
  \label{Prompt}
\end{figure}
 
The proposed FES-Net aims to address some of the difficulties encountered in retinal vessel segmentation by improving the original image and bypassing the need for conventional image processing methods. The network employs four convolutional blocks on the downsampling end and a relatively simpler upsampling end to generate an output binary mask for each category. To keep the overall computational cost low, only a limited number of transposed convolutions are used in the upsampling procedure. The final stage assigns a label to each vessel pixel using the pixel classification layer.

\section{Related Work}
\label{sec:Related Work}

In recent years, lightweight segmentation networks have gained attention in image segmentation challenges. SegNAS3D \cite{wong2019segnas3d} introduced a network search to optimize the 3D segmentation structure and significantly reduce the complexity of the model. RefineNet \cite{lin2017refinenet} was used as the lightweight backbone network by Nekrasov {\it et al.} \cite{nekrasov2018light} to improve the performance of the model. IC-Net \cite{zhao2018icnet} is a real-time semantic segmentation network that uses an image cascade with branch training to accelerate model convergence. BiSeNet \cite{yu2021bisenet} is a lightweight model based on a dual-branch topology, employing multiple paths to refine spatial information. DMFNet \cite{yuan2019dmfnet} partitions channels into numerous groups and employs a weighted three-dimensional expanded convolution. This approach ultimately results in a decrease in the parameter count while simultaneously enhancing the accuracy of inference. Xception \cite{chollet2017xception} and MobileNet \cite{howard2017mobilenets} use deep separable convolution to improve the inference speed. The Dense-Inception U-Net \cite{zhang2020dense} uses a compact encoder with a lightweight Inception backbone and a dense module to capture high-level semantic information. This architecture is designed for medical image segmentation tasks.  

ShuffleNet \cite{zhang2018shufflenet, ma2018shufflenet} applies group convolution and channel shuffling techniques to reduce computational expenses compared to more complex models. Iqbal {\it et al.} \cite{iqbal2022g} explored the development of lightweight segmentation networks specifically for medical images. However, creating networks with low model complexity, high inference speed, and excellent performance still poses a challenge in most medical image segmentation tasks. nnU-Net \cite{isensee2021nnu} improves network adaptability by preprocessing data and postprocessing segmentation results, but model parameters increase with this approach. A lightweight V-Net \cite{lei2020lightweight} uses convolution inverted residual bottleneck block (IRB block). To ensure segmentation accuracy and use fewer parameters, depth-wise convolution and point-wise convolution are employed. However, it does not speed up the inference process. Furthermore, Tarasiewicz {\it et al.} \cite{tarasiewicz2020lightweight} developed Lightweight U-Nets that can accurately delineate brain tumors from multimodal magnetic resonance imaging and trained several skinny networks in all image planes. PyConvU-Net \cite{li2021pyconvu} increases segmentation accuracy while using fewer parameters by replacing all traditional convolutional layers of U-Net with pyramidal convolution. However, its inference speed is low. G-Net Light \cite{iqbal2022g}, PLVS-Net \cite{arsalan2022prompt}, and MKIS-Net\cite{khan2022mkis} are effective CNN architectures to segment retinal blood vessels while being lightweight.

\section{Proposed Method}\label{method}
The conventional semantic segmentation networks are deep and have many trainable parameters, such as SegNet \cite{Badrinarayanan2017}, U-Net \cite{Ronneberger2015}, and DeepLab \cite{chen2018encoder}. These networks are based on encoder-decoder designs, where the decoder mirrors the encoder. This means that the network has double number of layers. Our proposed FES-Net, on the other hand, utilizes shallow upsampling that helps to reduce the depth of the network, resulting in parameter reduction. This network is based on four prompt convolutional blocks (PCBs), which are a combination of general convolutions and separable convolutions to reduce the cost of the network (Figs.~\ref{FES-NET} and \ref{Prompt}).


Each PCB in FES-Net comprises a 3×3 general convolution, a separable convolution of 3×3, and one 1×1 general convolution (Fig.~\ref{Prompt}). These convolutions are directly connected to a strided (dilated) convolution and employ batch normalization and rectified linear units (ReLU) for the combination of their outputs. The depth-wise concatenation layer is used to merge the outputs of these three convolutions, and FES-Net contains four such PCBs. The main feature of FES-Net is the enhancement of spatial features achieved through the feature enhancement block (FEB). FEB consists of a shallow convolutional block that preserves low-level features, does not significantly downsample them, and provides them at the end of the network. It uses four convolutions with a maximum depth of 16 channels only, which minimizes the network's parameters. FEB enhances spatial information on the initial layers of the network and provides rich features to the final stage of the network.

In the FES-Net architecture with FEB (Fig.~\ref{FES-NET}), the upsampling block receives the feature $F_D$ from the downsampling side, and from the shallow upsampling, it outputs the feature $F_{US}$, which is common in semantic segmentation networks. Continuous downsampling and multiple convolutional layers result in feature deterioration and $F_{US}$ cannot provide a better true positive rate. Therefore, FEB takes the characteristic $F_i$ from the initial layers of the network and generates the characteristic $F_E$ after the shallow convolutional operation, which contains low-level characteristics. The two characteristics $F_{US}$ and $F_E$ are combined to produce the characteristic $S_C$:
\begin{equation}\label{Eq1}
 S_C = F_{US}\,{\rm{\copyright}}\,F_{E}
\end{equation}
where the symbol ${\rm{\copyright}}$ denotes depth-wise concatenation of the two features, namely the feature on the upsampling side and the feature on FEB. The resulting feature $S_C$ contains rich edge information, resulting in higher sensitivity. Finally, at the end of the network, FES-Net uses the pixel classification layer to assign a predicted label to each pixel of the image.

\begin{table}[!t]
  \centering
  \caption{Performance comparison of the proposed FES-Net and several alternative methods on the DRIVE dataset.}
     \resizebox{1.0\textwidth}{!}{%
    \begin{tabular}{lccccc}
    \toprule
    \multirow{2}[4]{*}{\textbf{Method}} & \multicolumn{5}{c}{\textbf{Performance (\%)}} \\
\cmidrule{2-6} & \textbf{Se ($\uparrow$)} & \textbf{Sp ($\uparrow$)} & \textbf{Acc ($\uparrow$)} & \textbf{AUC ($\uparrow$)} & \textbf{F1 ($\uparrow$)} \\
    \midrule
    HED \cite{xie2015holistically} & 76.27 & 98.01 & 95.24 & 97.58 & 80.89 \\
    DRIU \cite{maninis2016deep} & -     & -     & -     & 97.93 & 82.20 \\
    DeepVessel \cite{fu2016deepvessel} & 76.12 & 97.68 & 95.23 & 97.52 & - \\
    Orlando et al. \cite{orlando2016discriminatively} & 78.97 & 96.84 & 94.54 & 95.06 & - \\
    JL-UNet \cite{yan2018joint} & 76.53 & 98.18 & 95.42 & 97.52 & - \\
    Att UNet \cite{oktay2018attention} & 79.46 & 97.89 & 95.64 & 97.99 & 82.32 \\
    H-DenseUNet \cite{li2018h} & 79.85 & 98.05 & 95.73 & 98.10 & 82.79 \\
    Three-stage CNN \cite{yan2018three} & 76.31 & 98.20 & 95.33 & 97.50 & - \\
    BTS-DSN \cite{Guo2019} & 78.00 & 98.06 & 95.51 & 97.96 & 82.08 \\
    DUNet \cite{Jin2019} & 79.84 & 98.03 & 95.75 & 98.11 & 82.49 \\
    M2U-Net \cite{laibacher2019m2u} & -     & -     & 96.30 & 97.14 & 80.91 \\
    Bio-Net \cite{xiang2020bio} & 82.20 & 98.04 & 96.09  & 98.26 & 82.06\\
    CC-Net \cite{Feng2020} & 76.25 & 98.09 & 95.28 & 96.78 & - \\
    CTF-Net \cite{wang2020ctf} & 78.49 & 98.13 & 95.67 & 97.88 & 82.41 \\
    CSU-Net \cite{wang2020csu} & 80.71 & 97.82 & 95.65 & 98.01 & 82.51 \\
    OCE-Net \cite{OCE-NET} & 80.18 & 98.26 & 95.81 & 98.21 & 83.02 \\
    Wave-Net \cite{liu2022wave} & 81.64 & 97.64 & 95.61 & -     & 82.54 \\
    LightEyes \cite{guo2022lighteyes} & -     & -     & -     & 97.96 & - \\
    G-Net Light \cite{iqbal2022g} & 81.92 & 98.29 & 96.86 & -     & 82.02 \\
    \midrule
    \textbf{Proposed FES-Net} & \textbf{82.89} & \textbf{98.31} & \textbf{96.91} & \textbf{98.32} & \textbf{83.10} \\
    \bottomrule
    \end{tabular}%
    }
  \label{DRIVE}%
\end{table}%

\begin{table}[!t]
  \centering
  \caption{Performance comparison of the proposed FES-Net and several alternative methods on the STARE dataset.}
   \resizebox{1.0\textwidth}{!}{%
    \begin{tabular}{lccccc}
    \toprule
    \multirow{2}[4]{*}{\textbf{Method}} & \multicolumn{5}{c}{\textbf{Performance (\%)}} \\
\cmidrule{2-6} & \textbf{Se ($\uparrow$)} & \textbf{Sp ($\uparrow$)} & \textbf{Acc ($\uparrow$)} & \textbf{AUC ($\uparrow$)} & \textbf{F1 ($\uparrow$)} \\
    \midrule
    HED \cite{xie2015holistically} & 80.76 & 98.22 & 96.41 & 98.24 & 82.68 \\
    Orlando et al. \cite{orlando2016discriminatively} & 76.80 & 97.38 & 95.19 & 95.70 & - \\
    JL-UNet \cite{yan2018joint} & 75.81 & 98.46 & 96.12 & 98.01 & - \\
    Att UNet \cite{oktay2018attention} & 77.09 & 98.48 & 96.33 & 97.00 & - \\
    H-DenseUNet \cite{li2018h} & 78.59 & 98.42 & 96.44 & 98.47 & 82.32 \\
    Three-stage CNN \cite{yan2018three} & 77.35 & 98.57 & 96.38 & 98.33 & - \\
    BTS-DSN \cite{Guo2019} & 82.01 & 98.28 & 96.60 & 98.72 & 83.62 \\
    DUNet \cite{Jin2019} & 78.92 & 98.16 & 96.34 & 98.43 & 82.30 \\
    CC-Net \cite{Feng2020} & 80.67 & 98.16 & 96.32 & 98.33 & 81.36 \\
    OCE-Net \cite{OCE-NET} & 80.12 & 98.65 & 96.72 & \textbf{98.76} & 83.41 \\
    Wave-Net \cite{liu2022wave} & 79.02 & 98.36 & 96.41 & -     & 81.40 \\
    LightEyes \cite{guo2022lighteyes} & -     & -     & -     & 98.29 & - \\
    G-Net Light \cite{iqbal2022g} & 81.70 & 98.53 & 97.30 & -     & 81.78 \\
    \midrule
    \textbf{Proposed FES-Net} & \textbf{82.97} & \textbf{98.91} & \textbf{97.33} & 97.17 & \textbf{83.99} \\
    \bottomrule
    \end{tabular}%
    }
  \label{STARE}%
\end{table}%

\begin{table}[!t]
  \centering
  \caption{Performance comparison of the proposed FES-Net and several alternative methods on the CHASE dataset.}
   \resizebox{1.0\textwidth}{!}{%
    \begin{tabular}{lccccc}
    \toprule
    \multirow{2}[4]{*}{\textbf{Method}} & \multicolumn{5}{c}{\textbf{Performance (\%)}} \\
\cmidrule{2-6} & \textbf{Se ($\uparrow$)} & \textbf{Sp ($\uparrow$)} & \textbf{Acc ($\uparrow$)} & \textbf{AUC ($\downarrow$)} & \textbf{F1 ($\uparrow$)} \\
    \midrule
    HED \cite{xie2015holistically} & 75.16 & 98.05 & 95.97 & 97.96 & 78.15 \\
    DeepVessel \cite{fu2016deepvessel} & 74.12 & 97.01 & 96.09 & 97.90 & - \\
    Orlando et al. \cite{orlando2016discriminatively} & 75.65 & 96.55 & 94.67 & 94.78 & - \\
    JL-UNet \cite{yan2018joint} & 76.33 & 98.09 & 96.10 & 97.81 & - \\
    Att UNet \cite{oktay2018attention} & 80.10 & 98.04 & 96.42 & 98.40 & 80.12 \\
    H-DenseUNet \cite{li2018h} & 78.93 & 97.92 & 96.11 & 98.35 & 79.01 \\
    Three-stage CNN \cite{yan2018three} & 76.41 & 98.06 & 96.07 & 97.76 & - \\
    BTS-DSN \cite{Guo2019} & 78.88 & 98.01 & 96.27 & 98.40 & 79.83 \\
    DUNet \cite{Jin2019} & 77.35 & 98.01 & 96.18 & 98.39 & 79.32 \\
    M2U-Net \cite{laibacher2019m2u} & -     & -     & 97.03 & 96.66 & 80.06 \\
    OCE-Net \cite{OCE-NET} & 81.38 & 98.24 & 96.78 & \textbf{98.72} & 81.96 \\
    Wave-Net \cite{liu2022wave} & 82.83 & 98.21 & 96.64 & -     & 83.49 \\
    LightEyes \cite{guo2022lighteyes} & -     & -     & -     & 98.20 & - \\
    G-Net Light \cite{iqbal2022g} & 82.10 & 98.38 & 97.26 & -     & 80.48 \\
    \midrule
    \textbf{Proposed FES-Net} & \textbf{85.95} & \textbf{98.88} & \textbf{97.55} & 98.61 & \textbf{84.88} \\
    \bottomrule
    \end{tabular}
    }
  \label{CHASEDB1}%
\end{table}%

\begin{table}[!t]
  \centering
  \caption{Performance comparison of the proposed FES-Net and several alternative methods on the HRF dataset.}
   \resizebox{1.0\textwidth}{!}{%
    \begin{tabular}{lccccc}
    \toprule
    \multirow{2}[4]{*}{\textbf{Model }} & \multicolumn{5}{c}{\textbf{Performance (\%)}} \\
\cmidrule{2-6} & \textbf{Se ($\uparrow$)} & \textbf{Sp ($\uparrow$)} & \textbf{Acc ($\uparrow$)} & \textbf{AUC ($\uparrow$)} & \textbf{F1 ($\uparrow$)} \\
    \midrule
    U-Net \cite{Ronneberger2015} & -     & -     & 95.87 & 83.05 & 72.39 \\
    CS-Net \cite{mou2019cs} & -     & -     & 95.66 & 82.32 & 71.04 \\
    SA-Unet \cite{guo2021sa} & -     & -     & 95.64 & 82.70  & 71.18 \\
    SCS-Net \cite{wu2021scs} & -     & -     & 95.65 & 81.80  & 70.66 \\
    CogSeg \cite{zhang2021collaborative} & -     & -     & 96.22 & 84.31 & 74.75 \\
    SuperVessel \cite{SuperVessel} & -     & -     & 96.54 & 85.06 & 76.74 \\
    \midrule
    \textbf{Proposed FES-Net} & \textbf{80.26} & \textbf{98.37} & \textbf{97.05} & \textbf{88.55} & \textbf{79.98} \\
    \bottomrule
    \end{tabular}%
    }
  \label{HRF}%
\end{table}%

\section{Results and Discussion}\label{experimentalResults}
The efficacy of the proposed FES-Net for segmenting retinal blood vessels was evaluated on publicly available datasets and compared with previously published state-of-the-art methods.

\subsection{Datasets}

\textbf{DRIVE:} 
The DRIVE \cite{DRIVEdata} dataset has 40 color retinal images divided into 20 training images and 20 testing images, with a resolution of $584\times 565$ pixels. It includes patients of different ages diagnosed with diabetes, with seven images representing the early stages of mild diabetic retinopathy. Manual pixel annotations are available for vessel and background segmentation in both training and testing images.

\textbf{STARE:} The STARE \cite{STAREDataset} dataset consists of 20 color fundus images with $35^\circ$ field of view (FOV) and a resolution of $700\times 605$ pixels. The ground truth is established using two manual segmentation options, and 10 images are used for training, while the remaining 10 are used for testing.

\textbf{CHASE:} The CHASE \cite{Fraz2012} dataset consists of 28 color images of 14 British schoolchildren. Each image has a resolution of $999\times 996$ pixels and is centered on the optical disc. Two manual segmentation maps are provided for ground truth. The dataset does not have a specific separation of training and testing sets. For our experiments, we used the initial 20 images for training and the final 8 images for testing.

\textbf{HRF:} The HRF dataset \cite{HRFDataset} includes 45 images, consisting of 15 images each from healthy individuals, patients with glaucoma, and patients with diabetic retinopathy. The images have a resolution of $3504\times2336$ pixels and a viewing angle of $60^\circ$. Ground truth segmentation was performed by a team of specialists for each image.

\begin{figure*}[!t]

    \includegraphics[width=\textwidth]{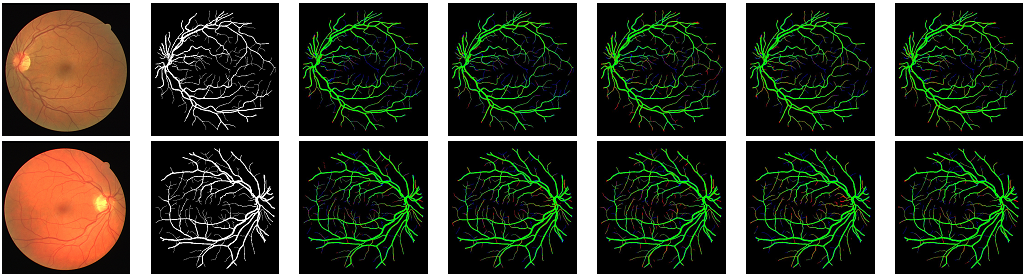} 
	\caption{Segmentation results of FES-Net and comparative methods on representative test images of the DRIVE dataset. From left to right: test image (\#1 in the top row and \#2 in the bottom row), ground truth segmentation, and the segmentation results of BCD-Unet, MultiResNet, SegNet, Unet++, and our proposed FES-Net, respectively. True positive pixels are shown in green, false positives in red, and false negatives in blue.}
	\label{visualDRIVE}%
\end{figure*}%

\begin{figure*}[!t]

    \includegraphics[width=\textwidth]{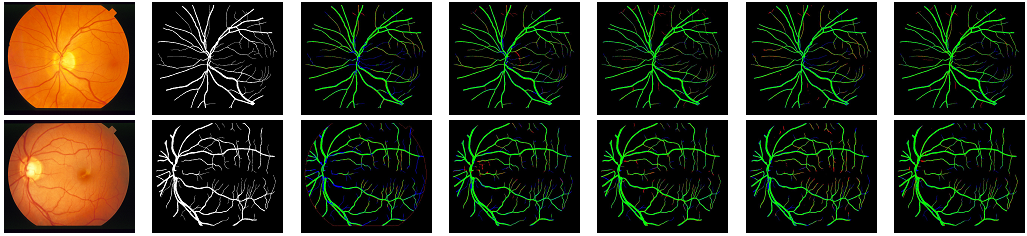} 
	\caption{Segmentation results of FES-Net and comparative methods on representative test images of the STARE dataset. From left to right: test image (\#2 in the top row and \#3 in the bottom row), ground truth segmentation, and the segmentation results of BCD-Unet, MultiResNet, SegNet, Unet++, and our proposed FES-Net, respectively. True positive pixels are shown in green, false positives in red, and false negatives in blue.}
 \label{visualSTARE}%
\end{figure*}%

\subsection{Experimental Setup}
For network training, the ADAM optimizer was used with a starting learning rate of 0.00002 and a decay rate of 0.90. Before starting the training, we resized the images for each dataset to a standard width of 640 pixels and normalized each image based on the z-score statistic. To enrich the dataset, we applied contrast enhancement, brightness adjustment, and random image flipping and rotation (ranging between 1--360$^\circ$). These techniques were employed to synthetically expand the number of training images.

\subsection{Evaluation Criteria}
Vessel segmentation maps exhibit a distinctive characteristic compared to other types of segmentation maps, as they adopt a binary representation, where each pixel is assigned exclusively to either the vessel or the background class. Accomplished ophthalmologists with specialized expertise manually annotated the ``ground truth'' labels in publicly accessible datasets. These annotations serve as a reference standard against which the performance of generated segmentation results can be assessed. The process involves the classification of individual image pixels into either the vascular or nonvascular category.
For each output image, there are four possible outcomes that can occur: true positive ($T_P$), which represents pixels correctly identified as vessels; true negative ($T_N$), which represents pixels correctly identified as non-vessels; false positive ($F_P$), which represents pixels mistakenly identified as vessels; and false negative ($F_N$), which represents pixels mistakenly identified as non-vessels. To allow comparisons between different methods, five widely used parameters are used in the field: sensitivity ($S_e$), specificity ($S_p$), accuracy ($A_{cc}$), area under the curve (\textit{AUC}), and the $F_1$ score. These measures assess different aspects of the performance of a method and their mathematical formulations are as follows:
\begin{equation}
S_{e}=\frac{T_{P}}{T_{P}+F_{N}},
\label{eq:Se}
\end{equation}
\begin{equation}
S_{p}=\frac{T_{N}}{T_{N}+F_{P}},
\label{eq:sp}
\end{equation}
\begin{equation}
A_{cc}=\frac{T_{P}+T_{N}}{T_{P}+T_{N}+F_{P}+F_{N}},
\label{eq:acc}
\end{equation}
\begin{equation}
\textit{AUC}=1-\frac{1}{2}\left(\frac{F_{P}}{F_{P}+T_{N}}+\frac{F_{N}}{F_{N}+T_{P}}\right),
\label{eq:AUC}
\end{equation}
\begin{equation}
F_{1} =\frac{2 \times T_{P}}{(2\times T_{P})+F_{P}+F_{N}}.
\label{eq:F1}
\end{equation} 

\subsection{Comparison and Experiments}
Here we present the results obtained with our proposed network, as well as a range of alternate approaches, on the DRIVE, STARE, CHASE, and HRF datasets. We commence by providing a prior qualitative and quantitative evaluation of a comprehensive spectrum of unsupervised and supervised techniques commonly employed in the field. This evaluation aims to establish a baseline performance and to compare the effectiveness of various existing methods. Subsequently, we perform an exhaustive comparison between U-Net \cite{Ronneberger2015} and SegNet \cite{Badrinarayanan2017}, which are widely recognized deep learning architectures for semantic segmentation tasks. The objective of this section is to present a self-contained comparison, while also showcasing the performance of our proposed network against these well-established benchmarks.

The quantitative results of our proposed FES-Net and other methods (Tables \ref{DRIVE}--\ref{HRF}) provide evidence that the proposed FES-Net consistently outperforms other techniques in terms of $A_{cc}$. Moreover, the results demonstrate that our proposed network obtains competitive performance in terms of $S_e$, $S_p$ and \textit{AUC}, particularly for the CHASE dataset, and consistently ranks among the top-performing models. In particular, there is no apparent discernible pattern among alternative methods in terms of optimal $S_e$, $S_p$ and \textit{AUC}, suggesting that the proposed FES-Net exhibits unique strengths in accurately segmenting retinal vessels across different datasets.

In addition to the quantitative analysis, we present a qualitative comparison of the results obtained by FES-Net and other methods on the DRIVE, STARE, and CHASE datasets. The results obtained on the DRIVE dataset (Fig.~\ref{visualDRIVE}) demonstrate that FES-Net significantly reduces false positives in small vessels compared to current methods. For example, U-Net variants struggle to accurately delineate vessel boundaries, resulting in a higher number of false positives, while SegNet \cite{Badrinarayanan2017} tends to generate false tiny vessels in most images, and BCD-Unet \cite{azad2019bi} appears to overlook crucial information about vessel structures, leading to suboptimal segmentation performance. In contrast, FES-Net effectively captures this information while minimizing the generation of false vessel information, resulting in more accurate segmentations.

Alternative methods tend to produce more false positives when applied to the STARE dataset (Fig.~\ref{visualSTARE}), particularly around the retinal boundaries, optic nerves, and small vessels. This may be attributed to the challenges posed by the complex retinal structures and image artifacts present in this dataset. The proposed FES-Net method proves to be more robust against these artifacts, preserving the fine details of the vessel structures while maintaining a low rate of false positives. These findings indicate that FES-Net is capable of capturing the subtle characteristics of retinal vessels in challenging scenarios, showcasing its efficacy in this domain.

\begin{figure*}[!t]
	\includegraphics[width=\textwidth]{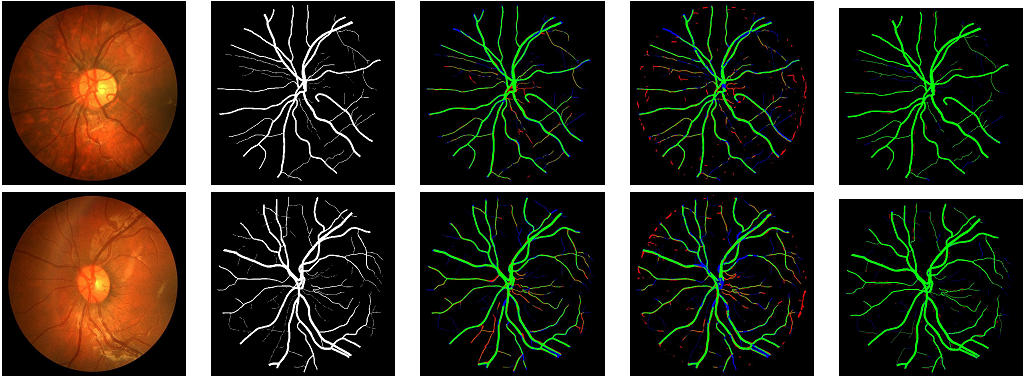}
		\caption{Segmentation results of FES-Net and comparative methods on representative test images of the CHASE dataset. From left to right: test image (\#1 in the top row and \#3 in the bottom row), ground truth segmentation, and the segmentation results of SegNet, VessSeg, and our proposed FES-Net, respectively. True positive pixels are shown in green, false positives in red, and false negatives in blue.}
	\label{visualCHASE}%
\end{figure*}%

Similar observations are made when applying FES-Net to the CHASE dataset (Fig.~\ref{visualCHASE}). Despite the presence of various challenges, such as image quality variations and vessel abnormalities, FES-Net consistently achieves accurate vessel segmentation. The proposed method effectively suppresses false positives while preserving the true vessel structures, even in regions with low contrast or overlapping vessel patterns. These results further establish the robustness and efficacy of FES-Net in a variety of datasets and image conditions.

In general, the comprehensive evaluation demonstrates the superiority of our proposed FES-Net method in terms of accuracy and robustness compared to the alternative methods examined. Quantitative analysis reveals consistent high performance, while visual comparisons highlight the ability of FES-Net to accurately capture retinal vessel structures with minimal false positives. These results provide solid evidence for the efficacy of our proposed method for retinal vessel segmentation tasks and its potential to help diagnose and monitor retinal diseases.

\subsection{Discussion}
There are notable architectural distinctions between our proposed network and the alternative models discussed above, which merit attention. First, it is important to emphasize that the FES-Net consistently outperforms the alternatives, with  a substantially reduced number of trainable parameters. Our network is designed to achieve a balance between performance and cost-effectiveness. Specifically, it consists of 1 million (M) trainable parameters, while for example, MultiResUNet \cite{IBTEHAZ202074} has about 7M and VessNet \cite{khan2020residual} about 9M parameters (Table~\ref{tab:timingComparison}).

In recent research, modifications to the U-Net and SegNet architectures have been introduced, leading to outstanding performance in the segmentation of retinal vessels \cite{azad2019bi,Zongwei2018}. However, it is crucial to emphasize that these methods often involve a substantially larger number of trainable parameters, typically an order of magnitude larger, in comparison to our proposed network. This suggests the potential of our approach to achieving comparable results with fewer parameters, which is promising from a computational perspective.

Furthermore, our experiments were conducted on publicly available standard datasets and performance metrics which are used widely. U-Net \cite{Ronneberger2015} and SegNet \cite{Badrinarayanan2017} are established benchmark methods in the field, and our proposed method consistently demonstrates strong competitiveness against the above-mentioned benchmarks, as evidenced by the results obtained by the FES-Net.

In general, our proposed network successfully balances performance and computational efficiency, delivering exceptional results with notably fewer trainable parameters compared to existing methods. This serves as compelling evidence for the effectiveness and potential of our approach for retinal vessel segmentation tasks.

\begin{table}[!t]
  \centering
  \caption{Computational requirements of the proposed FES-Net compared with several state-of-the-art methods.}
    \begin{tabular}{lcc}
    \toprule
    Method & Params ($\downarrow$) & Size ($\downarrow$) \\
     & (M) & (MB) \\
    \midrule
    MobileNet-V3-Small \cite{Howard2019MobileNet} & 2.5  & 11.0 \\
    ERFNet \cite{Romera2018ERFNet} & 2.1 & 8.0 \\
    MultiResUNet \cite{IBTEHAZ202074} & 7.2  & -- \\
    VessNet \cite{Arsalan2019} & 9.3  & 36.6  \\
    \midrule
    FES-Net & 1.0 & 3.6 \\
    \bottomrule
    \end{tabular}%
  \label{tab:timingComparison}%
\end{table}%

\section{Conclusions}
In this study, a novel feature enhancement network specifically designed for the segmentation of retinal blood vessels is proposed, taking into account the computational demands to deploy the network on resource-limited devices such as smartphones.
Significantly, our segmentation network achieves a remarkable decrease in the overall parameter count when compared to cutting-edge segmentation networks documented in the literature. It consists of approximately 1 million parameters only. We have presented extensive experiments and comparative analyses of different alternative methods using four publicly accessible datasets. The results provide ample evidence of the robustness, competitiveness, and effectiveness of our proposed network.


\begin{thebibliography}{10}
\providecommand{\url}[1]{#1}
\csname url@samestyle\endcsname
\providecommand{\newblock}{\relax}
\providecommand{\bibinfo}[2]{#2}
\providecommand{\BIBentrySTDinterwordspacing}{\spaceskip=0pt\relax}
\providecommand{\BIBentryALTinterwordstretchfactor}{4}
\providecommand{\BIBentryALTinterwordspacing}{\spaceskip=\fontdimen2\font plus
\BIBentryALTinterwordstretchfactor\fontdimen3\font minus
  \fontdimen4\font\relax}
\providecommand{\BIBforeignlanguage}[2]{{%
\expandafter\ifx\csname l@#1\endcsname\relax
\typeout{** WARNING: IEEEtran.bst: No hyphenation pattern has been}%
\typeout{** loaded for the language `#1'. Using the pattern for}%
\typeout{** the default language instead.}%
\else
\language=\csname l@#1\endcsname
\fi
#2}}
\providecommand{\BIBdecl}{\relax}
\BIBdecl

\bibitem{Fraz2012}
M.~M. {Fraz}, P.~{Remagnino}, A.~{Hoppe}, B.~{Uyyanonvara}, A.~R. {Rudnicka},
  C.~G. {Owen}, and S.~A. {Barman}, ``An ensemble classification-based approach
  applied to retinal blood vessel segmentation,'' \emph{IEEE Transactions on
  Biomedical Engineering}, vol.~59, no.~9, pp. 2538--2548, 2012.

\bibitem{imtiaz2021screening}
R.~Imtiaz, T.~M. Khan, S.~S. Naqvi, M.~Arsalan, and S.~J. Nawaz, ``Screening of
  glaucoma disease from retinal vessel images using semantic segmentation,''
  \emph{Computers \& Electrical Engineering}, vol.~91, p. 107036, 2021.

\bibitem{crosby2012retinal}
R.~Crosby-Nwaobi, L.~Z. Heng, and S.~Sivaprasad, ``Retinal vascular calibre,
  geometry and progression of diabetic retinopathy in type 2 diabetes
  mellitus,'' \emph{Ophthalmologica}, vol. 228, no.~2, pp. 84--92, 2012.

\bibitem{habib2014association}
M.~S. Habib, B.~Al-Diri, A.~Hunter, and D.~H. Steel, ``The association between
  retinal vascular geometry changes and diabetic retinopathy and their role in
  prediction of progression--an exploratory study,'' \emph{BMC Ophthalmology},
  vol.~14, no.~1, pp. 1--11, 2014.

\bibitem{naqvi2019automatic}
S.~S. Naqvi, N.~Fatima, T.~M. Khan, Z.~U. Rehman, and M.~A. Khan, ``Automatic
  optic disk detection and segmentation by variational active contour
  estimation in retinal fundus images,'' \emph{Signal, Image and Video
  Processing}, vol.~13, pp. 1191--1198, 2019.

\bibitem{naveed2021towards}
K.~Naveed, F.~Abdullah, H.~A. Madni, M.~A. Khan, T.~M. Khan, and S.~S. Naqvi,
  ``Towards automated eye diagnosis: An improved retinal vessel segmentation
  framework using ensemble block matching 3d filter,'' \emph{Diagnostics},
  vol.~11, no.~1, p. 114, 2021.

\bibitem{iqbal2022recent}
S.~Iqbal, T.~M. Khan, K.~Naveed, S.~S. Naqvi, and S.~J. Nawaz, ``Recent trends
  and advances in fundus image analysis: A review,'' \emph{Computers in Biology
  and Medicine}, vol. 151, p. 106277, 2022.

\bibitem{Zhao2018}
Y.~Zhao, J.~Xie, P.~Su, Y.~Zheng, Y.~Liu, J.~Cheng, and J.~Liu, ``Retinal
  artery and vein classification via dominant sets clustering-based vascular
  topology estimation,'' in \emph{Medical Image Computing and Computer Assisted
  Intervention (MICCAI)}, 2018, pp. 56--64.

\bibitem{khan2023retinal}
T.~M. Khan, S.~S. Naqvi, A.~Robles-Kelly, and I.~Razzak, ``Retinal vessel
  segmentation via a multi-resolution contextual network and adversarial
  learning,'' \emph{Neural Networks}, 2023.

\bibitem{iqbal2023robust}
S.~Iqbal, K.~Naveed, S.~S. Naqvi, A.~Naveed, and T.~M. Khan, ``Robust retinal
  blood vessel segmentation using a patch-based statistical adaptive
  multi-scale line detector,'' \emph{Digital Signal Processing}, p. 104075,
  2023.

\bibitem{iqbal2023ldmres}
S.~Iqbal, T.~M. Khan, M.~Alhussein, S.~S. Naqvi, M.~Usman, and K.~Aurangzeb,
  ``{LDMRes-Net}: Enabling real-time disease monitoring through efficient image
  segmentation,'' \emph{arXiv:2306.06145}, 2023.

\bibitem{Soomro2018}
T.~A. Soomro, T.~M. Khan, M.~A.~U. Khan, J.~Gao, M.~Paul, and L.~Zheng,
  ``Impact of {ICA}-based image enhancement technique on retinal blood vessels
  segmentation,'' \emph{IEEE Access}, vol.~6, pp. 3524--3538, 2018.

\bibitem{Khawaja2019a}
A.~{Khawaja}, T.~M. {Khan}, K.~{Naveed}, S.~S. {Naqvi}, N.~U. {Rehman}, and
  S.~{Junaid Nawaz}, ``An improved retinal vessel segmentation framework using
  {Frangi} filter coupled with the probabilistic patch based denoiser,''
  \emph{IEEE Access}, vol.~7, pp. 164\,344--164\,361, 2019.

\bibitem{khan2018robust}
K.~B. Khan, A.~A. Khaliq, A.~Jalil, and M.~Shahid, ``A robust technique based
  on {VLM} and {Frangi} filter for retinal vessel extraction and denoising,''
  \emph{PLoS One}, vol.~13, no.~2, p. e0192203, 2018.

\bibitem{Soomro2019}
T.~A. Soomro, A.~J. Afifi, J.~Gao, O.~Hellwich, L.~Zheng, and M.~Paul,
  ``Strided fully convolutional neural network for boosting the sensitivity of
  retinal blood vessels segmentation,'' \emph{Expert Systems with
  Applications}, vol. 134, pp. 36--52, 2019.

\bibitem{khan2020region}
T.~M. Khan, M.~Mehmood, S.~S. Naqvi, and M.~F.~U. Butt, ``A region growing and
  local adaptive thresholding-based optic disc detection,'' \emph{Plos one},
  vol.~15, no.~1, p. e0227566, 2020.

\bibitem{khan2022t}
T.~M. Khan, A.~Robles-Kelly, and S.~S. Naqvi, ``{T-Net}: A resource-constrained
  tiny convolutional neural network for medical image segmentation,'' in
  \emph{IEEE/CVF Winter Conference on Application of Computer Vision (WACV)},
  2022, pp. 644--653.

\bibitem{Ronneberger2015}
O.~Ronneberger, P.~Fischer, and T.~Brox, ``{U-Net}: Convolutional networks for
  biomedical image segmentation,'' in \emph{Medical Image Computing and
  Computer-Assisted Intervention (MICCAI)}, 2015, pp. 234--241.

\bibitem{Gu2019CENetCE}
Z.~Gu, J.~Cheng, H.~Fu, K.~Zhou, H.~Hao, Y.~Zhao, T.~Zhang, S.~Gao, and J.~Liu,
  ``{CE-Net}: Context encoder network for {2D} medical image segmentation,''
  \emph{IEEE Transactions on Medical Imaging}, vol.~38, pp. 2281--2292, 2019.

\bibitem{yan2018joint}
Z.~Yan, X.~Yang, and K.-T. Cheng, ``Joint segment-level and pixel-wise losses
  for deep learning based retinal vessel segmentation,'' \emph{IEEE
  Transactions on Biomedical Engineering}, vol.~65, no.~9, pp. 1912--1923,
  2018.

\bibitem{Wang2019a}
B.~Wang, S.~Qiu, and H.~He, ``Dual encoding {U-Net} for retinal
  vesselsegmentation,'' in \emph{Medical Image Computing and Computer Assisted
  Intervention (MICCAI)}, 2019, pp. 84--92.

\bibitem{fu2016deepvessel}
H.~Fu, Y.~Xu, S.~Lin, D.~W. Kee~Wong, and J.~Liu, ``{DeepVessel}: Retinal
  vessel segmentation via deep learning and conditional random field,'' in
  \emph{Medical Image Computing and Computer-Assisted Intervention (MICCAI)},
  2016, pp. 132--139.

\bibitem{wong2019segnas3d}
K.~C.~L. Wong and M.~Moradi, ``{SegNAS3D}: Network architecture search with
  derivative-free global optimization for {3D} image segmentation,'' in
  \emph{Medical Image Computing and Computer-Assisted Intervention (MICCAI)},
  2019, pp. 393--401.

\bibitem{lin2017refinenet}
G.~Lin, A.~Milan, C.~Shen, and I.~Reid, ``{RefineNet}: Multi-path refinement
  networks for high-resolution semantic segmentation,'' in \emph{IEEE
  Conference on Computer Vision and Pattern Recognition (CVPR)}, 2017, pp.
  1925--1934.

\bibitem{nekrasov2018light}
V.~Nekrasov, C.~Shen, and I.~Reid, ``Light-weight {RefineNet} for real-time
  semantic segmentation,'' \emph{arXiv:1810.03272}, 2018.

\bibitem{zhao2018icnet}
H.~Zhao, X.~Qi, X.~Shen, J.~Shi, and J.~Jia, ``{ICNet} for real-time semantic
  segmentation on high-resolution images,'' in \emph{European Conference on
  Computer Vision (ECCV)}, 2018, pp. 405--420.

\bibitem{yu2021bisenet}
C.~Yu, C.~Gao, J.~Wang, G.~Yu, C.~Shen, and N.~Sang, ``{BiSeNet V2}: Bilateral
  network with guided aggregation for real-time semantic segmentation,''
  \emph{International Journal of Computer Vision}, vol. 129, no.~11, pp.
  3051--3068, 2021.

\bibitem{yuan2019dmfnet}
J.~Yuan, W.~Zhou, and T.~Luo, ``{DMFNet}: Deep multi-modal fusion network for
  {RGB-D} indoor scene segmentation,'' \emph{IEEE Access}, vol.~7, pp.
  169\,350--169\,358, 2019.

\bibitem{chollet2017xception}
F.~Chollet, ``{Xception}: Deep learning with depthwise separable
  convolutions,'' in \emph{IEEE Conference on Computer Vision and Pattern
  Recognition (CVPR)}, 2017, pp. 1251--1258.

\bibitem{howard2017mobilenets}
A.~G. Howard, M.~Zhu, B.~Chen, D.~Kalenichenko, W.~Wang, T.~Weyand,
  M.~Andreetto, and H.~Adam, ``{MobileNets}: Efficient convolutional neural
  networks for mobile vision applications,'' \emph{arXiv:1704.04861}, 2017.

\bibitem{zhang2020dense}
Z.~Zhang, C.~Wu, S.~Coleman, and D.~Kerr, ``{DENSE-INception U-net for medical
  image segmentation},'' \emph{Computer Methods and Programs in Biomedicine},
  vol. 192, p. 105395, 2020.

\bibitem{zhang2018shufflenet}
X.~Zhang, X.~Zhou, M.~Lin, and J.~Sun, ``{ShuffleNet}: An extremely efficient
  convolutional neural network for mobile devices,'' in \emph{IEEE Conference
  on Computer Vision and Pattern Recognition (CVPR)}, 2018, pp. 6848--6856.

\bibitem{ma2018shufflenet}
N.~Ma, X.~Zhang, H.-T. Zheng, and J.~Sun, ``{ShuffleNet V2}: Practical
  guidelines for efficient {CNN} architecture design,'' in \emph{European
  Conference on Computer Vision (ECCV)}, 2018, pp. 116--131.

\bibitem{iqbal2022g}
S.~Iqbal, S.~S. Naqvi, H.~A. Khan, A.~Saadat, and T.~M. Khan, ``{G-Net Light}:
  A lightweight modified {Google Net} for retinal vessel segmentation,''
  \emph{Photonics}, vol.~9, no.~12, p. 923, 2022.

\bibitem{isensee2021nnu}
F.~Isensee, P.~F. Jaeger, S.~A. Kohl, J.~Petersen, and K.~H. Maier-Hein,
  ``{nnU-Net}: a self-configuring method for deep learning-based biomedical
  image segmentation,'' \emph{Nature Methods}, vol.~18, no.~2, pp. 203--211,
  2021.

\bibitem{lei2020lightweight}
T.~Lei, W.~Zhou, Y.~Zhang, R.~Wang, H.~Meng, and A.~K. Nandi, ``{Lightweight
  V-Net} for liver segmentation,'' in \emph{IEEE International Conference on
  Acoustics, Speech and Signal Processing (ICASSP)}, 2020, pp. 1379--1383.

\bibitem{tarasiewicz2020lightweight}
T.~Tarasiewicz, M.~Kawulok, and J.~Nalepa, ``{Lightweight U-Nets} for brain
  tumor segmentation,'' in \emph{International MICCAI Brain Lesion Workshop},
  2020, pp. 3--14.

\bibitem{li2021pyconvu}
C.~Li, Y.~Fan, and X.~Cai, ``{PyConvU-Net}: A lightweight and multiscale
  network for biomedical image segmentation,'' \emph{BMC Bioinformatics},
  vol.~22, no.~1, pp. 1--11, 2021.

\bibitem{arsalan2022prompt}
M.~Arsalan, T.~M. Khan, S.~S. Naqvi, M.~Nawaz, and I.~Razzak, ``Prompt deep
  light-weight vessel segmentation network {(PLVS-Net)},'' \emph{IEEE/ACM
  Transactions on Computational Biology and Bioinformatics}, vol.~20, no.~2,
  pp. 1363--1371, 2022.

\bibitem{khan2022mkis}
T.~M. Khan, M.~Arsalan, A.~Robles-Kelly, and E.~Meijering, ``{MKIS-Net}: A
  light-weight multi-kernel network for medical image segmentation,''
  \emph{arXiv:2210.08168}, 2022.

\bibitem{Badrinarayanan2017}
V.~Badrinarayanan, A.~Kendall, and R.~Cipolla, ``{SegNet}: A deep convolutional
  encoder-decoder architecture for image segmentation,'' \emph{IEEE
  Transactions on Pattern Analysis and Machine Intelligence}, vol.~39, no.~12,
  pp. 2481--2495, 2017.

\bibitem{chen2018encoder}
L.-C. Chen, Y.~Zhu, G.~Papandreou, F.~Schroff, and H.~Adam, ``Encoder-decoder
  with atrous separable convolution for semantic image segmentation,''
  \emph{Lecture Notes in Computer Science}, p. 833–851, 2018.

\bibitem{xie2015holistically}
S.~Xie and Z.~Tu, ``Holistically-nested edge detection,'' in \emph{IEEE
  International Conference on Computer Vision (ICCV)}, 2015, pp. 1395--1403.

\bibitem{maninis2016deep}
K.-K. Maninis, J.~Pont-Tuset, P.~Arbel{\'a}ez, and L.~V. Gool, ``Deep retinal
  image understanding,'' in \emph{Medical Image Computing and Computer-Assisted
  Intervention (MICCAI)}, 2016, pp. 140--148.

\bibitem{orlando2016discriminatively}
J.~I. Orlando, E.~Prokofyeva, and M.~B. Blaschko, ``A discriminatively trained
  fully connected conditional random field model for blood vessel segmentation
  in fundus images,'' \emph{IEEE Transactions on Biomedical Engineering},
  vol.~64, no.~1, pp. 16--27, 2016.

\bibitem{oktay2018attention}
O.~Oktay, J.~Schlemper, L.~L. Folgoc, M.~Lee, M.~Heinrich, K.~Misawa, K.~Mori,
  S.~McDonagh, N.~Y. Hammerla, B.~Kainz, B.~Glocker, and D.~Rueckert,
  ``{Attention U-Net}: Learning where to look for the pancreas,''
  \emph{arXiv:1804.03999}, 2018.

\bibitem{li2018h}
X.~Li, H.~Chen, X.~Qi, Q.~Dou, C.-W. Fu, and P.-A. Heng, ``{H-DenseUNet}:
  Hybrid densely connected {UNet} for liver and tumor segmentation from {CT}
  volumes,'' \emph{IEEE Transactions on Medical Imaging}, vol.~37, no.~12, pp.
  2663--2674, 2018.

\bibitem{yan2018three}
Z.~Yan, X.~Yang, and K.-T. Cheng, ``A three-stage deep learning model for
  accurate retinal vessel segmentation,'' \emph{IEEE Journal of Biomedical and
  Health Informatics}, vol.~23, no.~4, pp. 1427--1436, 2018.

\bibitem{Guo2019}
S.~Guo, K.~Wang, H.~Kang, Y.~Zhang, Y.~Gao, and T.~Li, ``{BTS-DSN}: Deeply
  supervised neural network with short connections for retinal vessel
  segmentation,'' \emph{International Journal of Medical Informatics}, vol.
  126, pp. 105--113, 2019.

\bibitem{Jin2019}
Q.~Jin, Z.~Meng, T.~D. Pham, Q.~Chen, L.~Wei, and R.~Su, ``{DUNet}: A
  deformable network for retinal vessel segmentation,'' \emph{Knowledge-Based
  Systems}, vol. 178, pp. 149--162, 2019.

\bibitem{laibacher2019m2u}
T.~Laibacher, T.~Weyde, and S.~Jalali, ``{M2U-Net}: Effective and efficient
  retinal vessel segmentation for resource-constrained environments,'' in
  \emph{IEEE/CVF Conference on Computer Vision and Pattern Recognition
  Workshops}, 2019, pp. 1--10.

\bibitem{xiang2020bio}
T.~Xiang, C.~Zhang, D.~Liu, Y.~Song, H.~Huang, and W.~Cai, ``{BiO-Net}:
  Learning recurrent bi-directional connections for encoder-decoder
  architecture,'' in \emph{Medical Image Computing and Computer-Assisted
  Intervention (MICCAI)}, 2020, pp. 74--84.

\bibitem{Feng2020}
S.~Feng, Z.~Zhuo, D.~Pan, and Q.~Tian, ``{CcNet}: A cross-connected
  convolutional network for segmenting retinal vessels using multi-scale
  features,'' \emph{Neurocomputing}, vol. 392, pp. 268--276, 2020.

\bibitem{wang2020ctf}
K.~Wang, X.~Zhang, S.~Huang, Q.~Wang, and F.~Chen, ``{CTF-Net}: Retinal vessel
  segmentation via deep coarse-to-fine supervision network,'' in \emph{IEEE
  International Symposium on Biomedical Imaging (ISBI)}, 2020, pp. 1237--1241.

\bibitem{wang2020csu}
B.~Wang, S.~Wang, S.~Qiu, W.~Wei, H.~Wang, and H.~He, ``{CSU-Net}: A context
  spatial {U-Net} for accurate blood vessel segmentation in fundus images,''
  \emph{IEEE Journal of Biomedical and Health Informatics}, vol.~25, no.~4, pp.
  1128--1138, 2020.

\bibitem{OCE-NET}
X.~Wei, K.~Yang, D.~Bzdok, and Y.~Li, ``Orientation and context entangled
  network for retinal vessel segmentation,'' \emph{arXiv:2207.11396}, 2022.

\bibitem{liu2022wave}
Y.~Liu, J.~Shen, L.~Yang, H.~Yu, and G.~Bian, ``{Wave-Net}: A lightweight deep
  network for retinal vessel segmentation from fundus images,'' \emph{Computers
  in Biology and Medicine}, vol. 152, p. 106341, 2022.

\bibitem{guo2022lighteyes}
S.~Guo, ``{LightEyes}: A lightweight fundus segmentation network for mobile
  edge computing,'' \emph{Sensors}, vol.~22, no.~9, p. 3112, 2022.

\bibitem{mou2019cs}
L.~Mou, Y.~Zhao, L.~Chen, J.~Cheng, Z.~Gu, H.~Hao, H.~Qi, Y.~Zheng, A.~Frangi,
  and J.~Liu, ``{CS-Net}: Channel and spatial attention network for curvilinear
  structure segmentation,'' in \emph{Medical Image Computing and
  Computer-Assisted Intervention (MICCAI)}, 2019, pp. 721--730.

\bibitem{guo2021sa}
C.~Guo, M.~Szemenyei, Y.~Yi, W.~Wang, B.~Chen, and C.~Fan, ``{SA-UNet}: Spatial
  attention {U-Net} for retinal vessel segmentation,'' in \emph{International
  Conference on Pattern Recognition (ICPR)}, 2021, pp. 1236--1242.

\bibitem{wu2021scs}
H.~Wu, W.~Wang, J.~Zhong, B.~Lei, Z.~Wen, and J.~Qin, ``{SCS-Net}: A scale and
  context sensitive network for retinal vessel segmentation,'' \emph{Medical
  Image Analysis}, vol.~70, p. 102025, 2021.

\bibitem{zhang2021collaborative}
Q.~Zhang, G.~Yang, and G.~Zhang, ``Collaborative network for super-resolution
  and semantic segmentation of remote sensing images,'' \emph{IEEE Transactions
  on Geoscience and Remote Sensing}, vol.~60, pp. 1--12, 2021.

\bibitem{SuperVessel}
Y.~Hu, Z.~Qiu, D.~Zeng, L.~Jiang, C.~Lin, and J.~Liu, ``{SuperVessel}:
  Segmenting high-resolution vessel from low-resolution retinal image,''
  \emph{arXiv:2207.13882}, 2022.

\bibitem{DRIVEdata}
J.~Staal, M.~D. Abr{\`a}moff, M.~Niemeijer, M.~A. Viergever, and
  B.~Van~Ginneken, ``{Ridge-based vessel segmentation in color images of the
  retina},'' \emph{IEEE Transactions Medical Imaging}, vol.~23, no.~4, pp.
  501--509, 2004.

\bibitem{STAREDataset}
A.~Hoover, V.~Kouznetsova, and M.~Goldbaum, ``{Locating blood vessels in
  retinal images by piecewise threshold probing of a matched filter
  response},'' \emph{IEEE Transactions Medical Imaging}, vol.~19, no.~3, pp.
  203--210, 2000.

\bibitem{HRFDataset}
J.~Odstrcilik, R.~Kolar, A.~Budai, J.~Hornegger, J.~Jan, J.~Gazarek, T.~Kubena,
  P.~Cernosek, O.~Svoboda, and E.~Angelopoulou, ``{Retinal vessel segmentation
  by improved matched filtering: evaluation on a new high-resolution fundus
  image database},'' \emph{IET Image Processing}, vol.~7, no.~4, pp. 373--383,
  2013.

\bibitem{azad2019bi}
R.~Azad, M.~Asadi-Aghbolaghi, M.~Fathy, and S.~Escalera, ``Bi-directional
  {ConvLSTM U-net} with densley connected convolutions,'' in \emph{IEEE
  International Conference on Computer Vision Workshops}, 2019.

\bibitem{IBTEHAZ202074}
N.~Ibtehaz and M.~S. Rahman, ``{MultiResUNet}: Rethinking the {U-Net}
  architecture for multimodal biomedical image segmentation,'' \emph{Neural
  Networks}, vol. 121, pp. 74--87, 2020.

\bibitem{khan2020residual}
T.~M. Khan, M.~Alhussein, K.~Aurangzeb, M.~Arsalan, S.~S. Naqvi, and S.~J.
  Nawaz, ``Residual connection-based encoder decoder network ({RCED-Net}) for
  retinal vessel segmentation,'' \emph{IEEE Access}, vol.~8, pp.
  131\,257--131\,272, 2020.

\bibitem{Zongwei2018}
Z.~Zhou, M.~M. Rahman~Siddiquee, N.~Tajbakhsh, and J.~Liang, ``{UNet++}: A
  nested {U-Net} architecture for medical image segmentation,'' in \emph{Deep
  Learning in Medical Image Analysis \& Multimodal Learning for Clinical
  Decision Support}, 2018, pp. 3--11.

\bibitem{Howard2019MobileNet}
A.~Howard, M.~Sandler, B.~Chen, W.~Wang, L.-C. Chen, M.~Tan, G.~Chu,
  V.~Vasudevan, Y.~Zhu, R.~Pang, H.~Adam, and Q.~Le, ``Searching for
  {MobileNetV3},'' in \emph{IEEE/CVF International Conference on Computer
  Vision (ICCV)}, 2019, pp. 1314--1324.

\bibitem{Romera2018ERFNet}
E.~Romera, J.~M. Álvarez, L.~M. Bergasa, and R.~Arroyo, ``{ERFNet}: Efficient
  residual factorized convnet for real-time semantic segmentation,'' \emph{IEEE
  Transactions on Intelligent Transportation Systems}, vol.~19, no.~1, pp.
  263--272, 2018.

\bibitem{Arsalan2019}
M.~Arsalan, M.~Oqais, tahir Mahmood, S.~W. Cho, and K.~R. Park, ``Aiding the
  diagnosis of diabetic and hypertensive retinopathy using artificial
  intelligence-based semantic segmentation,'' \emph{Journal of Clinical
  Medicine}, vol.~8, no.~9, 2019.

\end{thebibliography}
\end{document}